\documentclass[12pt]{article}
\usepackage[cp1251]{inputenc}  
\usepackage[english]{babel}
\usepackage{amsmath,amssymb,epsfig,graphicx,doc,latexsym,amsfonts,bm,longtable}
\usepackage[bookmarksopen,colorlinks]{hyperref}
\usepackage[monochrome]{color}
 \hypersetup{colorlinks=true}

\begin{document}
\begin{center}
{\textbf{\LARGE{Ion-kinetic D'Angelo mode}}}

\bigskip
{\large{D.V. Chibisov}},{\large{V.S. Mikhailenko} and K.N. Stepanov}

\bigskip

Kharkov National University, 61108 Kharkov, Ukraine

\end{center}

\begin{abstract}
An extension of hydrodynamic D'Angelo mode of inhomogeneous sheared
plasma flow along the magnetic field into the short-wavelength
limit, where the hydrodynamic treatment is not valid, has been
considered. We find that D'Angelo mode in this wavelength range is
excited by inverse ion Landau damping and becomes the shear flow
driven ion-kinetic mode.

\end{abstract}

\section{INTRODUCTION}

It is well established now that existence of magnetic field aligned
plasma sheared flows is a fundamental reality throughout the space
plasmas \cite{Amatucci,Ergun,Koepke}. The presence of velocity shear
produces a drastic changing in the stability properties of these
plasma flows, development of specific shear flow driven
instabilities and plasma turbulence, which is responsible for
efficient mechanisms for ion energization. A substantial amount of
research has been devoted to hydrodynamic electrostatic instability
of such shear flows, which was originally discussed by D'Angelo
\cite{DA65} and named now as a D'Angelo mode. This instability is of
Kelvin-Helmholtz type. It was investigated experimentally and
theoretically in different conditions specifically both in
collisionless \cite{DA66} and collisional \cite{Willig} plasmas, in
a plasmas with negative ions \cite{An,DA91} as well as in dusty
plasmas \cite{DA90,DA01}. In homogeneous plasmas D'Angelo mode is a
purely growing with zero frequency and growth rate
$\gamma=k_{z}v_{s}\left(\left(k_{y}/k_{z}\right)S-1\right)^{1/2}$,
where $v_{s}=\left(T_{e}/m_{i}\right)^{1/2}$ is the ion sound
velocity, $k_{y}$ and $k_{z}$ are the components  of the wave vector
across and along the magnetic field respectively,
$S_{i}=\left(1/\omega_{ci}\right)\left(dV_{0}\left(X\right)/dX\right)$
is the normalized ion flow velocity shear, $\omega_{ci}$ is the ion
cyclotron frequency. Instability develops, when
$\left(k_{y}/k_{z}\right)S>1$. This mode has obvious difference with
shear flow modified ion sound instability \cite{Gav}, which requires
that $\left(k_{y}/k_{z}\right)S<1$ and develops due to inverse
electron and ion Landau damping. In Ref.\cite{Mikhailenko07}
Mikhailenko et al. by using the kinetic approach, which accounted
for the effects of the finite Larmor radius, electron and ion Landau
damping, have developed the comprehensive theory of the hydrodynamic
and kinetic shear flow modified and shear flow driven low frequency
ion-sound-drift instabilities in inhomogeneous plasma shear flow
under condition $\left(k_{y}/k_{z}\right)S<1$. The purpose of this
paper is to present a kinetic theory of the electrostatic low
frequency instabilities of inhomogeneous plasmas under condition
$\left(k_{y}/k_{z}\right)S>1$, which is key to connection of the
investigated instabilities with D'Angelo mode. In contrary to
conclusions of Smith and Goeler \cite{Smith}, which have considered
this instability on the base of the Vlasov equation and
climed\cite{Smith} that ion Landau damping leads only to
modification of this hydrodynamical instability, we find that in
short wavelength limit, where hydrodynamic treatment is not valid at
all, the ion-kinetic effects are dominant and are responsible for
the development of new kinetic instability. Keeping in mind that
this instability is characterizes by positive parameter
$\left(k_{y}/k_{z}\right)S$, we nominate it as a kinetic D'Angelo
mode. This kinetic D'Angelo mode occurs when velocity shear exceeds
the critical value, which equals to one for the development of the
hydrodynamic D'Angelo mode \cite{DA65} and develops due to inverse
ion Landau damping. We find, that for the velocity shear less or of
the order of the critical value the kinetic D'Angelo mode transforms
into the drift instability, which has a similar mechanism of
excitation.

The paper is organized as follows. In Sec. 2 the dispersion relation for D'Angelo mode
obtained by using  the kinetic approach is represented. In Sec. 3 the hydrodynamic
D'Angelo mode for inhomogeneous plasma shortly reviewed with
accounting for the effects of finite ion Larmor radius. Sec. 4 is devoted to analysis
of the kinetic D'Angelo mode in the short wavelength limit.
In Sec. 5 the drift instability when the velocity shear less of
order the critical value is considered. Also here a numerical
analysis is presented. Conclusions are given in Sec. 6.

\section{DISPERSION RELATION FOR LOW-FREQUENCY INSTABILITIES DRIVEN BY SHEAR FLOW }

We consider the inhomogeneous sheared plasma flow with one ion
species which moves with velocity
$\textbf{V}_{0}\left(x\right)\parallel B_{0}\textbf{e}_{z}$. The
general dispersion relation in the kinetic Vlasov-Poisson model was
obtained in Ref.\cite{Mikhailenko06}
\begin{eqnarray}
\varepsilon\left({\mathbf{k},\omega
}\right)=1+\frac{1}{k^2\lambda_{De}^2}\left(1+i\sqrt{\pi}z_{e\ast}
W\left(z_{e0}\right)\right)+\frac{1}{k^2\lambda_{Di}^2}
\left[1-\frac{k_{y}}{k_{z}}S_{i} + i\sqrt{\pi}\sum\limits_{n
=-\infty}^{\infty}W\left(z_{i n}\right)
A_{n}\left(b_{i}\right)\right.\nonumber\\
\left.\times\left(z_{i\ast}
-\frac{k_{y}}{k_{z}}S_{i}z_{in}\right)\right]=0,\label{1}
\end{eqnarray}
where $\lambda_{D\alpha}$ is the Debye length, $\omega_{c\alpha}$
and $v_{T\alpha}=\sqrt{T_{\alpha}/m_{\alpha}}$ are the cyclotron
frequency and thermal velocity respectively,
$A_{n}\left(b_{i}\right)=I_{n}\left(b_{i}\right)e^{-b_{i}}$,
 $I_{n}$ is the modified Bessel function, $b_{i}= k_\bot^2 \rho_{Ti}^2$,
$\rho_{Ti}= v_{Ti}/\omega_{ci}$  is the thermal Larmor radius,
$z_{\alpha n} = \left(\omega-n\omega_{c\alpha} -
k_{z}V_{0\alpha}\right)/\sqrt{2}k_{z}v_{T\alpha}$, $z_{\alpha \ast}
= \left(\omega-k_{z}V_{0\alpha}-k_{y}v_{d\alpha}
\right)/\sqrt{2}k_{z}v_{T\alpha}$,
$v_{d\alpha}=\left(v_{T\alpha}^{2}/\omega_{c\alpha}\right)\left(d\ln
n_{0}\left(x\right)/dx\right)$ is the diamagnetic drift velocity,
and
$W\left(z\right)=\exp\left(-z^{2}\right)\left(1+\left(2i/\sqrt{\pi}
\right)\int\limits_{0}^{z}\exp\left(t^{2}\right)dt\right)$. In Eq.
(\ref{1}) is assumed that the local approximation, $k_{x}L_{n}\gg1$
and $k_{x}L_{v}\gg1$, where $L_{n}=\left[d\ln n_{0}\left(x\right)/dx
\right]^{-1}$, $L_{v}=\left[d\ln V_{0}\left(x\right)/dx
\right]^{-1}$, holds.

We consider the low frequency instability when the condition
$\omega=\omega\left(\mathbf{k}\right)+k_{z}V_{0}\ll\omega_{ci}$ is
met, so the zero term in the sum over cyclotron harmonics in Eq.
(\ref{1}) is the dominant.  We suppose that electrons are
adiabatical with $|z_{e0}|\ll 1$, and $|z_{i0}|$ may be greater or
less than unity, whereas $|z_{in}|>1$ for $n\neq0$. The last
condition is valid, when inequality $k_{z}\rho_{Ti}<1$ is satisfied.
Using the asymptotic form for $W$ - function for large argument
values, $W(z_{i})\thicksim
\exp\left(-z_{i}^{2}\right)+\left(i/\sqrt{\pi}z_{i}\right)\left(1+1/2z_{i}^{2}\right)$,
we carry out the summation over all cyclotron harmonics with
$n\neq0$. Then dispersion relation (\ref{1}) reduces to the form
\begin{eqnarray}
&\displaystyle k^2\lambda_{Di}^2\varepsilon\left({\mathbf{k},\omega
}\right)=1+\tau\nonumber \\
&\displaystyle+i\sqrt{\pi}W\left(z_{i0}\right)A_{0}\left(b_{i}
\right) \left(z_{i0}-\lambda k_{y}\rho_{Ti}S_{i}z_{i0}-\lambda
k_{y}\rho_{Ti}\frac{v_{di}}{\sqrt{2}v_{Ti}}\right)-\lambda
k_{y}\rho_{Ti}S_{i} A_{0}\left(b_{i} \right)=0,\label{2}
\end{eqnarray}
where $\tau=T_{i}/T_{e}$. In Eq. (\ref{2}) the variables
$\lambda=1/k_{z}\rho_{Ti}$ and
$z_{i0}=\left(\omega-k_{z}V_{0\alpha}\right)/\sqrt{2}k_{z}v_{T\alpha}$,
which are the normalized wavelength along the magnetic field and
complex frequency respectively, are used.  Eq. (\ref{2}) is the
general dispersion relation for the low frequency perturbations
which takes into account effects of the velocity shear, thermal
motion of ions, both along and across the magnetic field, as well as
the plasma density inhomogeneity. The analytical solution to Eq.
(\ref{2}) in general case for arbitrary $\lambda$ and $z_{i0}$
values can not be found. However,  it admits an approximate solution
in asymptotic limits of long as well as short wavelengths along the
magnetic field, when the effect of ion Landau damping is negligible
or dominant, respectively. In the long wavelength limit, this
equation gives hydrodynamic shear-flow-driven instability, known as
D'Angelo mode. In the short wavelength range, the source of free
energy for instability is also a shear velocity, however the
excitation of instability occurs via the inverse Landau damping and
therefore this instability is nominated as the ion kinetic D'Angelo
mode.

\section{HYDRODYNAMIC D'ANGELO MODE}

In the long wavelength limit, when inequality
$\left|z_{i0}\right|>1$ is met and ion Landau damping is negligible,
the dispersion relation (\ref{2}) reduces to the form
\begin{eqnarray}
2z_{i0}^{2}\left(1+\tau -A_{0}\left(b_{i}\right)\right)+\sqrt{2}
z_{i0}A_{0}\left(b_{i}\right)\frac{k_{y}v_{di}}{\omega_{ci}}\lambda
+A_{0}\left(b_{i}\right)\left(\lambda k_{y}\rho_{Ti}
S_{i}-1\right)=0.\label{3}
\end{eqnarray}
This equation gives the classical hydrodynamic D'Angelo mode
\cite{DA65}, which, in addition, accounts for the thermal motion of
ions across the magnetic field. The frequency and growth rate of
this mode respectively are
\begin{eqnarray}
\omega \left(k\right)=\frac{\omega_{0}}{2},\qquad
\gamma=\frac{\omega _{0} }{2} \sqrt{4\frac{\omega _{ci}^{2}
\left(1+\tau -A_{0} \left(b_{i} \right)\right)} {k_{y}^{2}
v_{di}^{2} A_{0} \left(b_{i} \right)\lambda ^{2} } \left(\lambda
k_{y} \rho _{Ti} S_{i} -1\right)-1},\label{4}
\end{eqnarray}
where
\begin{eqnarray}
\omega_{0}=-\frac{k_{y}v_{di}A_{0}\left(b_{i}\right)}{\left(1+\tau
-A_{0} \left(b_{i}\right)\right)}\label{5}
\end{eqnarray}
is the drift wave frequency. It follows from Eq.(\ref{4}), that the
hydrodynamic D'Angelo mode occurs in the wavelengths range
$\lambda_{2}<\lambda<\lambda_{1}$, where
\begin{eqnarray}
\lambda _{1,2} =\frac{2\omega _{ci}^{2} A_{0} \left(b_{i}
\right)k_{y} \rho _{Ti} S_{i} \pm \sqrt{\left(2\omega _{ci}^{2}
A_{0} \left(b_{i} \right)k_{y} \rho _{Ti} S_{i} \right)^{2} -\omega
_{0}^{2} 4\omega _{ci}^{2} \left(1+\tau -A_{0} \left(b_{i}
\right)\right)A_{0} \left(b_{i} \right)} }{\omega _{0}^{2}
\left(1+\tau -A_{0} \left(b_{i} \right)\right)}.\label{6}
\end{eqnarray}
The hydrodynamic D'Angelo mode exists when $\lambda_{1}$ and
$\lambda_{2}$ are both real and positive. The condition of the
wavelength positivity is satisfied when inequality
$k_{y}\rho_{Ti}S_{i}>0$ is met, whereas the wavelengths are real
when the flow velocity shear satisfies inequality
\begin{eqnarray}
S_{i} >\frac{\left|v_{di}\right| }{v_{Ti} }
\sqrt{\frac{A_{0}\left(b_{i}
\right)}{\left(1+\tau-A_{0}\left(b_{i}\right)\right)}}.\label{7}
\end{eqnarray}
In the limiting case $k_{\perp}\rho_{Ti}\ll1$ and $\tau=1$,
inequality (\ref{7}) reduces to the condition
$S_{i}>\left|v_{di}\right|/v_{Ti}$ obtained by D'Angelo in the fluid
treatment \cite{DA65}. Under condition (\ref{7}) the growth rate
(\ref{4}) has a maximum at $\lambda_{0}=2/k_{y}\rho_{Ti}S_{i}$.

The condition for validation of hydrodynamic treatment,
$\left|z_{i0} \right|>1$, at the extremities of interval
$\left(\lambda_{2};\lambda_{1}\right)$, as well as in the point of
the growth rate maximum, $\lambda_{0}$, should be verified also.  At
the long wavelength boundary,
\begin{eqnarray}
\lambda=\lambda_{1}\approx4S_{i} \frac{v_{Ti}^{2} }{v_{di}^{2} }
\frac{\left(1+\tau -A_{0} \left(b_{i} \right)\right)} {k_{y} \rho
_{Ti} A_{0} \left(b_{i} \right)},\label{8}
\end{eqnarray}
the modulus of  normalized frequency is approximately equals
\begin{eqnarray}
\left|z_{i0}\right|\approx\sqrt{2}S_{i}\frac{v_{Ti}
}{\left|v_{di}\right| }\label{9}
\end{eqnarray}
and hydrodynamic treatment is met at
$S_{i}>|v_{di}|/\sqrt{2}v_{Ti}$. This inequality is the less strict
than (\ref{7}) and therefore it holds if inequality (\ref{7}) is
satisfied. In the point of the growth rate maximum,  the modulus of
normalized frequency is
\begin{eqnarray}
\left|z_{i0}\right|\approx \frac{\left|v_{di}\right|
}{v_{Ti}S_{i}}\frac{A_{0}\left(b_{i}
\right)}{\left(1+\tau-A_{0}\left(b_{i}\right)\right)}\label{10}
\end{eqnarray}
and under condition (\ref{7}) inequality $\left|z_{i0} \right|>1$ is
not met here. This condition at $\lambda=\lambda_{2}$ is not
satisfied also. Hence the problem of the D'Angelo mode at the short
wavelength threshold, as well as in the maximum of the growth rate
can not be solved in the hydrodynamic treatment.

\section{KINETIC D'ANGELO MODE}

In order to investigate the D'Angelo mode in the short wave limit,
when Landau damping is dominant, we consider Eq. (\ref{2}) in
general form without appealing to the asymptotic forms of $W$
function. Find first the short wavelength threshold of the D'Angelo
mode. The threshold values for variables $\lambda$ and $z_{i0}$ are
determined by the balance of the velocity shear, plasma
inhomogeneity as well as thermal motion effects in ion Landau
damping and may be obtained by equating to zero the real and
imaginary parts of Eq. (\ref{2}),
\begin{eqnarray}
\left\{\begin{array}{l} {z_{i0} -\lambda k_{y} \rho _{Ti} S_{i}
z_{i0} -\lambda k_{y}v_{di}\left/\sqrt{2}\omega_{ci}\right.=0,} \\
{1+\tau -\lambda k_{y} \rho _{Ti} S_{i} A_{0} \left(b_{i}
\right)=0,} \end{array}\right.\label{11}
\end{eqnarray}
System of equations (\ref{11}) has a solution, when inequality
$k_{y}\rho_{Ti}S_{i}>0$ is met. The short-wavelength threshold value
$\lambda_{s}$ for the excitation of the instability, as well as the
threshold value of the normalized complex frequency $z_{s}$, which
is the real at that threshold, are
\begin{eqnarray}
\lambda_{s} =\frac{1+\tau }{k_{y} \rho _{Ti} S_{i} A_{0} \left(b_{i}
\right)} ,\label{12}
\end{eqnarray}
\begin{eqnarray}
z_{s}=-\frac{v_{di}}{\sqrt{2}S_{i}v_{Ti}}\frac{1+\tau}{1+\tau-A_{0}
\left(b_{i}\right)},\label{13}
\end{eqnarray}
where index $s$ means the short-wavelength instability threshold.
The approximate solution to Eq.(\ref{2}) in the vicinity of
instability threshold, we obtain by Taylor series expansion of
Eq.(\ref{3}) in powers of $\left(\lambda-\lambda_{s}\right)$ with
retained only zero-order and linear terms,
\begin{eqnarray}
z_{i0}\simeq
z_{s}+z'_{\lambda}\left(\lambda_{s}\right)\left(\lambda-\lambda_{s}\right).\label{14}
\end{eqnarray}
Here
$z'_{\lambda}\left(\lambda_{s}\right)=-\varepsilon'_{\lambda}/\varepsilon'_{z}$
and $\varepsilon'_{\lambda}$ and $\varepsilon'_{z}$ are determined
by
\begin{eqnarray}
k^{2}\lambda_{Di}^{2}\varepsilon'_{\lambda}
\left(\lambda_{s}\right)=-i\sqrt{\pi}W\left(z_{s}\right)A_{0}\left(b_{i}\right)\left(k_{y}\rho_{Ti}S_{i}
z_{s}+\frac{k_{y}v_{di}}{\sqrt{2}\omega_{ci}}\right)-k_{y}\rho_{Ti}S_{i}A_{0}\left(b_{i}\right)\label{15}
\end{eqnarray}
\begin{eqnarray}
k^{2}\lambda_{Di}^{2}\varepsilon'_{z
}\left(\lambda_{s}\right)=-i\sqrt{\pi}W\left(z_{s}\right)\left(1+\tau-A_{0}\left(b_{i}\right)\right).\label{16}
\end{eqnarray}
The expression (\ref{14}) does not claim to be exact solution of
dispersion relation, however it will give the ability to determine
the condition on the velocity shear at which the D'Angelo mode in
short wavelength range occurs. The  waves frequency and the growth
rate in the vicinity of the instability threshold can be obtained
from Eq. (\ref{14}) as
\begin{eqnarray}
\omega\left(k\right)\simeq\omega_{0}\frac{\lambda_{s}}{\lambda}-\frac{\omega_{0}
A_{0}\left(b_{i}\right)}{\left(1+\tau-A_{0}
\left(b_{i}\right)\right)}\left(1-\frac{\lambda_{s}}{\lambda}\right)+\omega_{ci}\frac{\sqrt{2}k_{y}\rho_{Ti}
S_{i} A_{0}\left(b_{i}\right)\text{Im}W\left(z_{s}\right)}{\sqrt{\pi
}\left|W\left(z_{s}
\right)\right|^{2}\left(1+\tau-A_{0}\left(b_{i}\right)\right)}\left(1-\frac{\lambda_{s}}{\lambda}\right),\label{17}
\end{eqnarray}
\begin{eqnarray}
\gamma\simeq\omega_{ci}\frac{\sqrt{2}k_{y}\rho_{Ti}S_{i}A_{0}\left(b_{i}\right)\text{Re}W\left(z_{s}\right)}{\sqrt{\pi}\left|W\left(z_{s}
\right)\right|^{2}\left(1+\tau-A_{0}\left(b_{i}\right)\right)}\left(1-\frac{\lambda_{s}}{\lambda}\right),\label{18}
\end{eqnarray}
As follows from Eq. (\ref{18}) the D'Angelo mode is a stable for the
wavelengths along the magnetic field which are satisfied to
inequality $\lambda\leq\lambda_{s}$, whereas for waves with
$\lambda>\lambda_{s}$ it becomes unstable. The instability is
induced due to effect of inverse ion Landau damping which occurs
when the free energy of velocity shear exceeds the  absorption of
energy by ion thermal motion along the magnetic field and plasma
inhomogeneity. Formally the instability exists for arbitrary
magnitudes of the velocity shear for wavelengths
$\lambda>\lambda_{s}$. However because
$\gamma\propto\text{Re}W\left(z_{s}\right)=
\exp\left(-z_{s}^{2}\right)$, the instability growth rate (\ref{18})
is exponentially small in the vicinity of the threshold when
$\left|z_{s}\right|>1$  and it is not exponentially small when
$\left|z_{s}\right|<1$. Lust inequality gives in fact the necessary
condition on normalized shear, at which the D'Angelo mode is
unstable. Accounting for Eq.(\ref{13}) yields
\begin{eqnarray}
S_{i}>S_{i0}=\frac{1+\tau
}{1+\tau-A_{0}\left(b_{i}\right)}\frac{\left|v_{di}\right|}{\sqrt{2}
v_{Ti}}.\label{19}
\end{eqnarray}
In the $k_{y}\rho_{Ti}\ll1$ limit and for $\tau=1$ inequality
(\ref{19}) reduces to $S_{i}>\sqrt{2}\left|v_{di}\right|/v_{Ti}$
which coincides with that obtained by D'Angelo in the fluid
treatment \cite{DA65}. The magnitude of the growth rate is affected
also by factor
$k_{y}\rho_{Ti}A_{0}\left(b_{i}\right)/(1+\tau-A_{0}\left(b_{i}\right))$
which gives the maximum of the growth rate at
$k_{y}\rho_{Ti}\simeq0.5$.

This kinetic instability reveals the main properties of the
hydrodynamic D'Angelo mode, such as an increase in growth rate with
increasing of velocity shear and it decreases with increasing of
plasma inhomogeneity. The latter property, however, is manifested
only in the quadratic term of the expansion (\ref{14}), which is
omitted here because of its lengthy form. It is interesting to note,
that the boundary wavelength (\ref{12}) for the kinetic mode is
independent on the plasma inhomogeneity, in distinction from
hydrodynamic mode. Thus, the D'Angelo mode in the vicinity of the
short wave threshold $\lambda_{s}$ occurs due to inverse ion Landau
damping which is caused by the parallel velocity shear and it is the
ion-kinetic D'Angelo mode. The necessary condition on velocity shear
value is expressed by Eq. (\ref{19}). The most unstable waves are
those with transverse wave numbers $k_{y}\rho_{Ti}\simeq0.5$.

The growth rate of the ion-kinetic D'Angelo mode (\ref{18})
increases with an increase of wavelength until condition
$\left|z_{i0}\right|<1$ holds. With subsequent increases of the
wavelength, so that inequality $\left|z_{i0}\right|>1$ is met, the
role of kinetic effects on the development of the D'Angelo mode
reduces, whereas the hydrodynamic mechanism becomes dominant and
ion-kinetic D'Angelo mode transformes gradually into the classical
hydrodynamic D'Angelo mode. In the transition wavelength region,
where $\left|z_{i0}\right|\simeq1$ and both the kinetic and
hydrodynamic effects are significant, an analytic solution of Eq.
(\ref{2}) not obtained. The corresponding range of the wavelengths
along the magnetic field may be studied only numerically.

\section{ION-KINETIC SHEAR-FLOW-DRIVEN DRIFT INSTABILITY}

When the velocity shear satisfies inequality $S_{i}\lesssim S_{i0}$,
so that $z_{s}\gtrsim1$, the D'Angelo mode does not exist, whereas
the kinetic growth rate, though small compared to D'Angelo mode, but
is not equal to zero. In the asymptotic limit
$z_{i0}>z_{s}\gtrsim1$, the dispersion relation (\ref{2}) for
inhomogeneous plasmas has a form
\begin{eqnarray}
2z_{i0}^{2}\left(1+\tau -A_{0}\left(b_{i}\right)\right)+\sqrt{2} z_{i0}
A_{0}\left(b_{i}\right)\frac{k_{y}V_{di}}{\omega_{ci}}\lambda
+A_{0}\left(b_{i}\right)\left(\lambda k_{y}\rho_{Ti} S_{i}-1\right)\nonumber\\
+2i\sqrt{\pi } A_{0} \left(b_{i} \right)z_{i0}^{2}
\left(z_{i0}-\lambda k_{y}\rho_{Ti}S_{i}z_{i0}-\lambda
k_{y}\rho_{Ti}\frac{v_{di}}{\sqrt{2}v_{Ti}}\right)\exp \left(-z_{i0}^{2} \right)=0, \label{20}
\end{eqnarray}
which is almost a same as Eq. (\ref{3}), but we retain in the
(\ref{20}) the ion Landau damping term. The frequency of
instability,
\begin{eqnarray}
\omega \left(k\right)\approx \omega_{0} +k_{z}v_{Ti}\frac{v_{Ti}
}{v_{di}}S_{i},\label{21}
\end{eqnarray}
corresponds to the drift waves modified by velocity shear. Therefore
this waves may be identified as a drift waves. The growth rate of
the drift instability,
\begin{eqnarray}
\gamma \approx\sqrt{\pi } \frac{\omega ^{2} \left(k\right)}{\sqrt{2}
\omega _{ci} } \left(k_{y} \rho _{Ti} S_{i} \lambda -\frac{1+\tau
}{A_{0} \left(b_{i} \right)} \right)\frac{A_{0} \left(b_{i}
\right)\lambda }{\left(1+\tau -A_{0} \left(b_{i} \right)\right)}
\exp \left(-\frac{\lambda ^{2} \omega ^{2}\left(k\right) }{2\omega
_{ci}^{2} } \right) \label{22}
\end{eqnarray}
is positive for wavelengths $\lambda>\lambda_{s}$, so, the drift
instability, as well as the ion kinetic d'Angelo mode, is excited
due to inverse ion Landau damping. The maximum of the growth rate
occurs approximately at wavelength
\begin{eqnarray}
\lambda_{0}\approx\frac{\lambda_{s}
}{2}\left(1+\sqrt{1+\frac{2S_{i}^{2}}{S_{i0}^{2}}}\right).\label{23}
\end{eqnarray}
With increasing wavelengths at $\lambda>\lambda_{0}$, the growth
rate decreases exponentially.

\begin{figure}
\includegraphics[width=12cm]{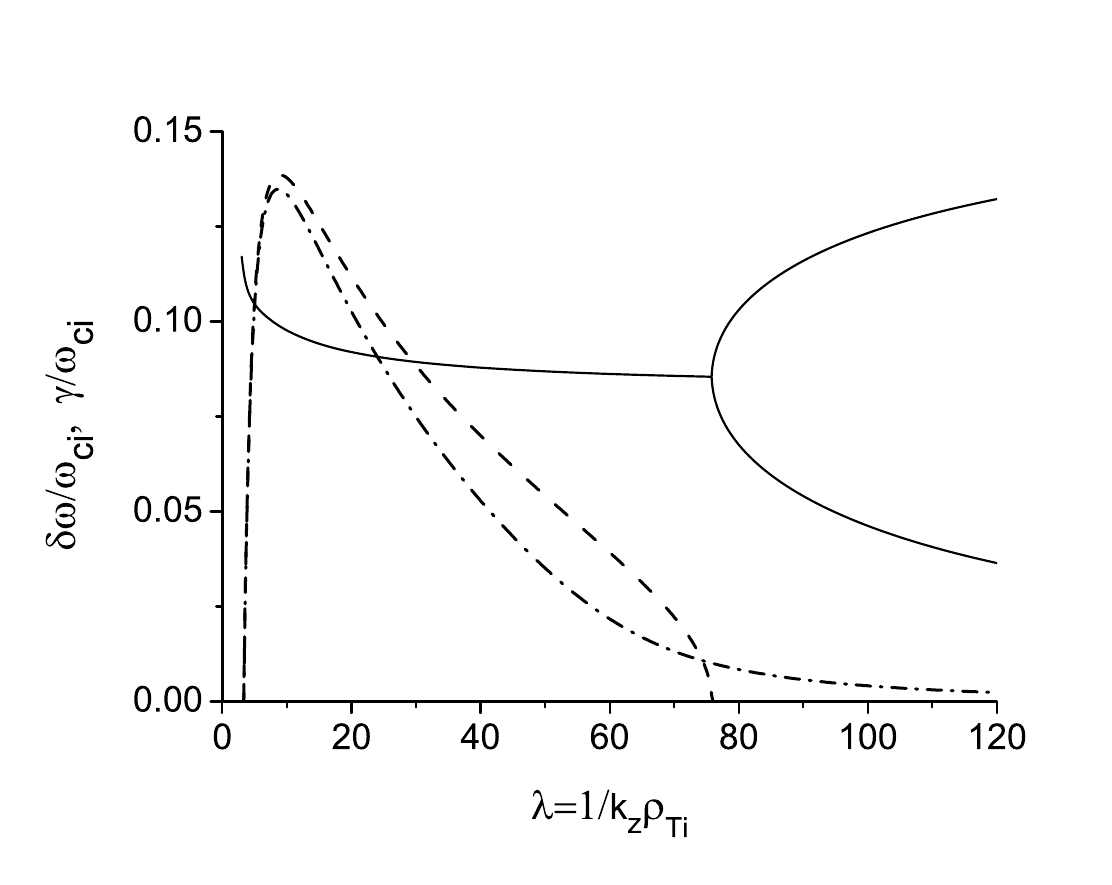}
\caption{The growth rate of the D'Angelo mode vs the normalized
wavelength: (1) $S_{i}=0.3$, (2) $S_{i}=0.2$, (3) $S_{i}=0.1$. Here
$|v_{di}|/v_{Ti}=0.1$, $k_{y}\rho_{Ti}=0.5$, $\tau=1$.}\label{Fig.1}
\end{figure}

In order to verify the results of analytical calculations, we solved
numerically the dispersion relation (\ref{1}) in the frequency range
$\omega\ll\omega_{ci}$. Figure \ref{Fig.1} shows the growth rate of
instability versus the normalized
wavelength for the different magnitudes of normalized velocity
shear, $S_{i}=0.3$,  $S_{i}=0.2$, and $S_{i}=0.1$ (curves 1, 2 and
3), which appropriate to ionospheric parameters \cite{Amatucci}.
Here we take also $|v_{di}|/v_{Ti}=0.1$, $k_{y}\rho_{Ti}=0.5$,
$\tau=1$. The condition for velocity shear (\ref{19}) is satisfied
only for the first two values of shear, so that only curves 1 and 2
present the D'Angelo mode, whereas curve 3 presents the drift
instability. For these conditions, the value of $|z_{s}|$ at the
short wavelength boundary is equal to 0.4, 0.6 and 1.2,
respectively. The short-wavelength boundary $\lambda_{s}$ calculated
from Eq. (\ref{13}) is equal approximately to 17, 25 and 51,
respectively, which agrees with Figure data. The long-wavelength
boundary $\lambda_{1}$ of the hydrodynamic D'Angelo mode for
$S_{i}=0.3$ and $S_{i}=0.2$ was calculated from Eq. (\ref{9}) and is
equal to 360 and 235, respectively. The growth rate for curves 1 and
2 has a maximum at wavelengths corresponding to
$\left|z_{i0}\right|\simeq1$ i. e. in the transient
kinetic-hydrodynamic range of the longitudinal wavelengths, where
the kinetic and hydrodynamic effects are significant both. For shear
value $S_{i}=0.1$ (curve 3), the condition of hydrodynamic approach
(\ref{8}) does not satisfied, and thus D'Angelo mode does not
develop. So this mode is the kinetic shear-flow driven drift
instability. The point of the maximal growth rate for this
instability, calculated from Eq. (\ref{23}), is approximately equal
$\lambda\approx65$, which is slight less than the value from the
Figure.

\section{CONCLUSIONS}
An extension of hydrodynamic D'Angelo mode of inhomogeneous sheared
plasma flow along the magnetic field into the short-wavelength
limit, where the hydrodynamic treatment is not valid, has been
considered. We find that D'Angelo mode in this wavelength range is
excited by inverse ion Landau damping and becomes the shear flow
driven ion-kinetic mode. The short-wavelength boundary for the wave
number component along the magnetic field $\lambda_{s}$ (\ref{12})
of instability development is determined by the balance of shear
flow, plasma inhomogeneity and thermal motion effects in ion Landau
damping. The kinetic D'Angelo mode is induced for the wavelengths
$\lambda>\lambda_{s}$ and velocity shear $S>S_{i0}$, where
$S_{i0}=\left(\left|v_{di}\right|/\sqrt{2}
v_{Ti}\right)\left(1+\tau\right)
/\left(1+\tau-A_{0}\left(b_{i}\right)\right)$ is the critical value.
This condition for velocity shear is the same as  for the ordinary
D'Angelo mode \cite{DA65}, so that kinetic and hydrodynamic D'Angelo
modes exist simultaneously both in different wavelength regions. The
kinetic D'Angelo mode reveals the main properties of the
hydrodynamic D'Angelo mode, such as an increase in growth rate with
increasing of velocity shear; and it decreases with increasing of
plasma inhomogeneity. However, the short wavelength boundary
(\ref{11}) for the kinetic mode appears to be independent on the
plasma inhomogeneity unlike the long wavelength boundary (\ref{8})
of hydrodynamic mode. The maximum of the growth rate of D'Angelo
mode occurs in transient wavelength range, where both the kinetic
and hydrodynamic effects are significant. For velocity shear value
$S\lesssim S_{i0}$ and wavelengths $\lambda<\lambda_{s}$ the
D'Angelo mode does not exist, however, the ion-kinetic drift
instability at $\lambda>\lambda_{s}$ is developed due to velocity
shear. This instability occurs as the kinetic D'Angelo mode, when
the free energy of velocity shear exceeds the absorption of energy
by ion thermal motion along the magnetic field and plasma
inhomogeneity. Formally, this instability exists for arbitrary
magnitudes of the velocity shear for wavelengths
$\lambda>\lambda_{s}$, however its growth rate (\ref{19}) is
exponentially small, when $\left|z_{s}\right|\gg1$.

\end{document}